\documentclass[a4]{article}
\usepackage{graphicx}

\begin{document}

\author{B. Brzostowski $^{1}$, M.R. Dudek $^{1}$ \footnote{ author: e-mail: {\sf mdudek@proton.if.uz.zgora.pl}},\\B. Grabiec $^{1}$ and  T. Nadzieja $^{2}$\\
$^{1}$ Institute of Physics, University of Zielona G{\'o}ra, \\65-069 Zielona G{\'o}ra, Poland\\
$^{2}$ Faculty of Mathematics, Computer Science and Econometrics, 
\\University of Zielona G{\'o}ra, 65-069 Zielona G{\'o}ra, Poland}   

\title{Non-finite-difference algorithm for  integrating Newton's motion equations }
\maketitle

\begin{abstract}
We have presented some practical consequences on the molecular-dynamics simulations arising from the numerical algorithm published recently in paper \cite{l_DudekNadzieja}. The algorithm is not a finite-difference method and therefore 
it could be  complementary to the traditional  numerical integrating of the motion equations. It consists of two steps. First, an analytic form of polynomials in some formal parameter $\lambda$ (we put $\lambda=1$ after all) is derived, which approximate the solution of the system of differential equations under consideration. Next, the numerical values of the derived polynomials  in the interval, in which the difference between them and their truncated part of smaller degree  does not exceed a given accuracy $\varepsilon$, become the numerical solution. 
The particular examples, which we have considered, represent the forced linear and nonlinear oscillator and the 2D Lennard-Jones fluid. In the latter case we have restricted to the polynomials of the first degree in formal parameter $\lambda$.

The computer simulations play very important role in modeling
materials with unusual properties being contradictictory to our intuition.
The particular example could be the auxetic materials. 
In this case, the accuracy of the applied numerical algorithms as well as 
various side-effects, which might change the physical reality, could become 
important for the properties of the simulated material. 
\end{abstract}

\section{Introduction}
Recently, we have published a numerical algorithm for the Cauchy problem for the ordinary differential equations \cite{l_DudekNadzieja}. We showed that it could be much more accurate, even by few orders of magnitude,  than traditional numerical methods based on finite differences. 
In physical applications, the requirement of one force evaluation per time step
makes that the most often chosen algorithm is the Verlet algorithm  \cite{l_Verlet,l_Berendsen}, being the simple third order 
Taylor predictor method, or  the equivalent   {\it leap-frog} algorithm \cite{l_leapfrog1,l_Berendsen}. In this case, the possibility to use algorithm being much more accurate then Verlet algorithm and as fast as the Verlet algorithm makes new perspective for simulating such complex systems as, e.g., tetratic phases \cite{l_Wojciechowski} or auxetics \cite{l_Lakes}--\cite{l_Konyok}. 
Apart from the problem of numerical accuracy there is also the possibility 
of the loss of the time-reversibility in  finite-difference methods \cite{l_HooverBook}, \cite{l_Toxvaerd}.

In the following, we discuss our algorithm with respect to integrating the motion equations. To this aim we have introduced a few examples of the forced linear and nonlinear oscillators and 2D Lennard-Jones fluid.

\begin{figure}
\begin{center}
\includegraphics[scale=0.3]{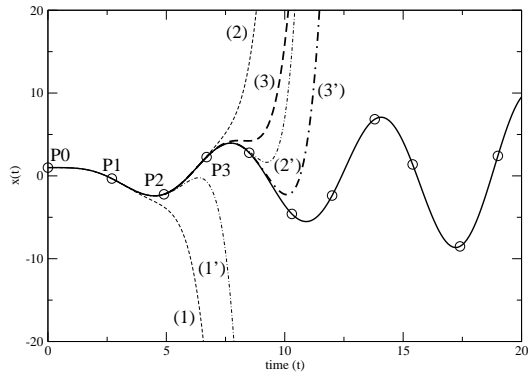}
\end{center}
\caption{The exact solution, $x(t)=\frac{1}{2}\,t\,\sin(t)+\cos(t)$, of the forced harmonic oscillator, $ x^{\prime\prime} + x = \cos(t)$ where $x_0=1, v_0=0$, (the thick line) and three pairs of approximating expressions $x_N$, (1) and (1$^{\prime}$) starting at point (P$_0$), (2) and (2$^{\prime}$) starting at point (P$_1$), (3) and (3$^{\prime}$) starting
at point (P$_2$). We have chosen $N=5$ in the approximations (1),(2),(3), whereas
$N=4$ in (1$^{\prime}$), (2$^{\prime}$), (3$^{\prime}$). In this example the number of exact digits is equal to 6, $\varepsilon=10^{-4}$.}
\label{fig1}
\end{figure}

\section{Short description of  the algorithm}
We present the procedure \cite{l_DudekNadzieja} of finding an approximate solution of the following initial value problem for the second order differential equation of the form: 

\begin{equation}
x^{\prime \prime} = f(x,v) + g(t),
\label{r_1}
\end{equation}

\begin{equation}
x(t_0)=x_0,\,\,\,\, x^{\prime}(t_0) =v_0,
\label{r_2}
\end{equation}

\noindent
where $x^{\prime}=v$, $f$ and $g$ are given functions, and $x_0$, $v_0$ are fixed reals.
For the function $f$ we assume that it is sufficiently smooth, so we can write $f$, using the Taylor formula, in some neighborhood of $(x_0, v_0)$ in the form

\begin{equation}
f(x,v) = \Sigma_{k=0}^N [(x-x_0) \frac{\partial}{\partial x} + (v-v_0)\frac{\partial}{\partial v}]^kf(x_0, v_0) + o([\sqrt{(x-x_0)^2 + (v-v_0)^2}]^N).
\label{r_3}
\end{equation}

\noindent
We introduce a formal real parameter $\lambda$ and instead of the Eqs.~(\ref{r_1}-\ref{r_2})  we consider the family of problems

\begin{equation}
x^{\prime \prime} = \lambda(f(x,v) + g(t)) 
\label{r_4}
\end{equation}

\noindent
with the initial data in Eq.~(\ref{r_2}). Next, we seek the approximate solution of Eq.~(\ref{r_4}) in the form

\begin{equation}
x_N(t) =x_0 +v_0 \tau +\Sigma_{k=1}^N \lambda ^k \varphi_k(\tau),
\label{r_5}
\end{equation}

\noindent
where $\varphi_k(\tau)$ are unknown functions of $\tau=t-t_0$ satisfying the condition
 
\begin{equation}
\varphi_k(0)=\varphi_k^{\prime}(0) = 0.
\label{r_6}
\end{equation}

\noindent
Putting Eq.(\ref{r_5}) into Eq.(\ref{r_4}) and next comparing the coefficients of order $\lambda ^k$ we get the system of differential equations for $\varphi_k$, which, together with initial conditions Eq.(\ref{r_6}),  determine $\varphi_k$ in the unique way. The differential equations for $\varphi_k$ we solve by simple integration.

\begin{figure}
\begin{center}
\includegraphics[scale=0.3]{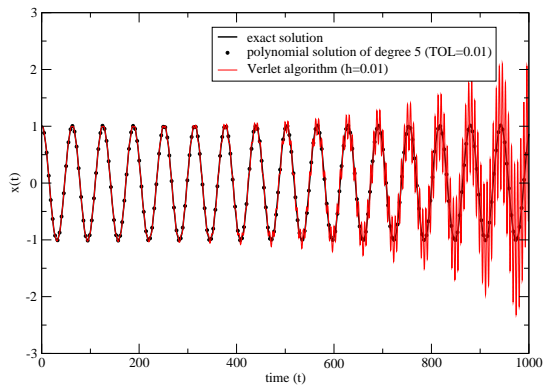}
\end{center}
\caption{The exact solution of the forced harmonic oscillator, $ x^{\prime \prime} + x = \cos(\omega t)$, (the thick line), the approximate solution generated by the Velocity-Verlet algorithm with step size $h=0.01$ and the solution generated by our polynomial algorithm (of order of $\lambda^5$). In this example $\varepsilon=10^{-2}$ and $\omega=0.1$.}
\label{fig2}
\end{figure}

To illustrate this procedure we consider the mathematical pendulum problem with external force
$\cos (t)$

\begin{equation}
x^{\prime \prime}= -x +\cos (t),\,\,\,x(t_0)=x_0,\,\,\,\,x^{\prime}(t_0) =v_0.
\end{equation}

\noindent
For $N=3$ we get

\begin{equation}
x_3 =x_0+ v_0\,\tau + \Sigma_{k=1} ^3 \lambda^k \varphi_k (\tau),
\end{equation}

\begin{equation}
x_3^{\prime \prime}=  \Sigma_{k=1} ^3 \lambda^k \varphi_k^{\prime \prime}(\tau) =-\lambda\,(x_0 + v_0\,\tau) - \Sigma_{k=1} ^3 \lambda^{k+1} \varphi_k(\tau) +\lambda \cos(t_0+\tau), 
\end{equation}

\noindent
where we substitute $t_0+\tau$ for $t$ and the derivatives with respect to $t$
 for the derivatives with respect to $\tau$. Hence,  

\begin{equation}
\varphi^{\prime \prime}_1(\tau) = -x_0 - v_0\,\tau+\cos (t_0+\tau),\,\,\,\,\varphi^{\prime \prime}_2(\tau) =-\varphi_1(\tau),\,\,\,\,\varphi^{\prime \prime}_3(\tau)= -\varphi_2(\tau),
\end{equation}

\noindent
and after integrating the above equations in the interval $[0,\tau]$ we obtain 

\begin{eqnarray}
\varphi_1(\tau) &=&\cos(t_0)-\tau\,\sin(t_0) - \cos(t_0+\tau) - \frac{\tau^3v_0}{6} - \frac{\tau^2 x_0}{2},\\
 \varphi_2(\tau)&=&\cos(t_0) -\tau\,\sin(t_0) - \cos(t_0+\tau) - \frac{\tau^2 \cos(t_0)}{2} \nonumber \\
&& +\frac{\tau^3 \sin(t_0)}{6} +\frac{\tau^5 v_0}{120} +\frac{\tau^4 x_0}{24},\\
\varphi_3(\tau) &=&\cos(t_0) - \tau\,\sin(t_0) - \cos(t_0 + \tau) -\frac{\tau^2\,\cos(t_0)}{2} \nonumber \\
&&+\frac{\tau^4\cos(t_0)}{24} +\frac{\tau^3\sin(t_0)}{6} -\frac{\tau^5\sin(t_0)}{120} - \frac{\tau^7v_0}{5040} -\frac{\tau^6 x_0}{720} 
\end{eqnarray}

We claim that for sufficiently large $N$ and $\lambda=1$ the expression $x_N(t)$ is a good approximation of the solution of the Eqs.~(\ref{r_1},\ref{r_2}) 
on a small interval of $t \in [t_0, t_0 +\delta_1]$.

Practically, for a fixed $N$ we look for the interval of $t \in [t_0, t_0 +\delta_1]$ such that 
\begin{equation}
|x_N (t) - x_{N-1} (t)|< \varepsilon,
\label{r_TOL}
\end{equation}

\noindent
 where $\varepsilon >0$ is a fixed accuracy. 
In the above example of mathematical pendulum the condition states that $| \varphi_3(t-t_0)|< \varepsilon$.
 Next, we repeat our procedure for the Eq.~(\ref{r_1}) with the new initial data 

\begin{equation}
x(t_0+\delta_1)=x_N (t_0 +\delta_1),\,\,\,\,\,x^{\prime}(t_0+\delta_1)=x^{\prime}_N (t_0 +\delta_1),
\end{equation}

\noindent
and so on. Fig.\ref{fig1} is a visualization of the updating procedure for the initial data. Thus, every time  the condition in Eq.(\ref{r_TOL}) fails at some value of $t_1>t_0$, the new initial data $\{t_0, x_0, v_0\}$ are defined  at $t_1$, i.e., $t_0=t_1,\,\, x_0=x(t_1),\,\, v_0=v(t_1)$.

In many examples it is enough to put $N=3$ to get the good approximation of the solution.
\begin{figure}
\begin{center}
\includegraphics[scale=0.3]{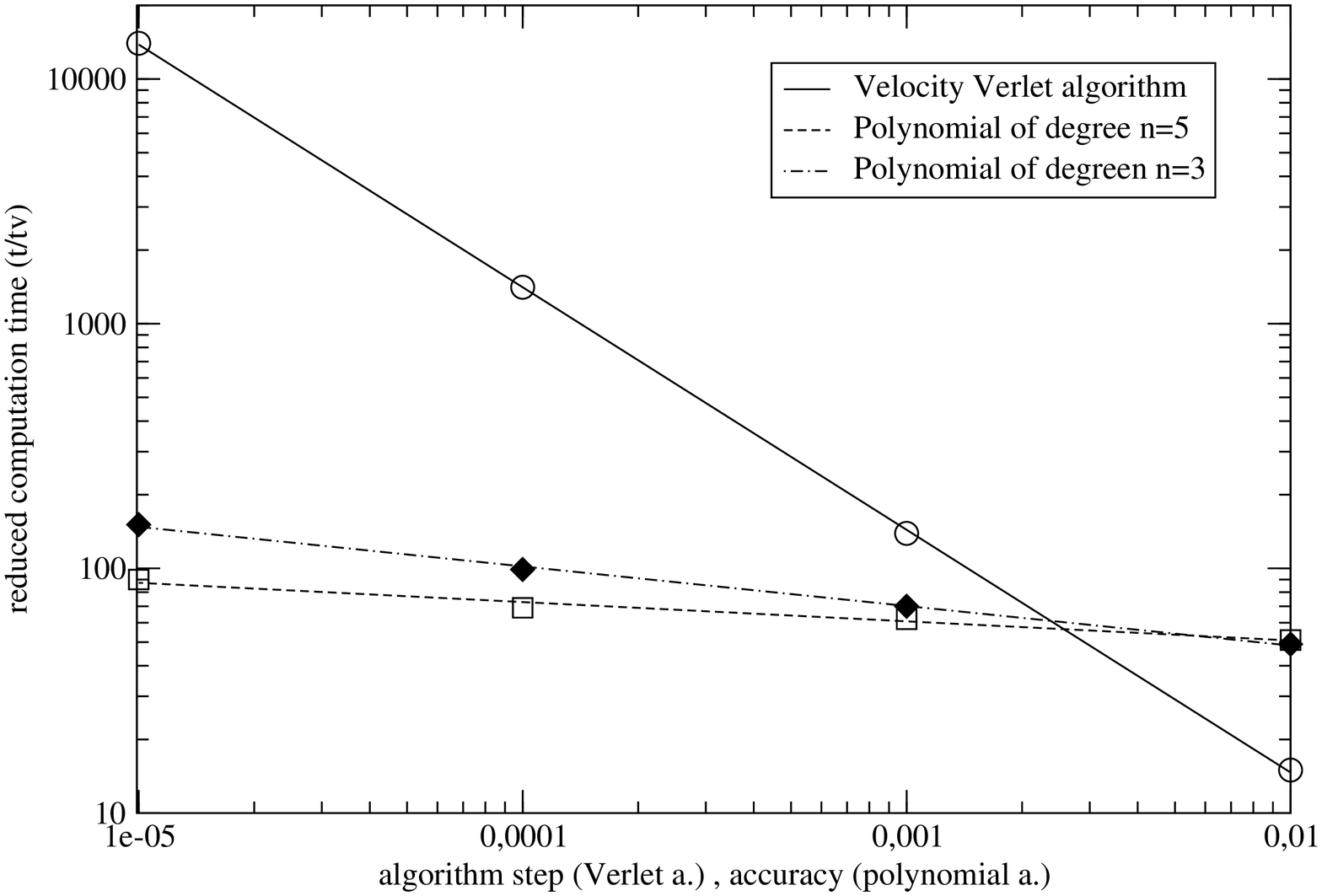}
\end{center}
\caption{Dependence of the total calculation time of the Velocity-Verlet algorithm and 
polynomial method on the chosen value $h$ (in case of the Verlet a.)  and the given accuracy $\varepsilon$ (in case of polynomial method). Time units, $t_v$, represent  calculation time of the Velocit-Verlet algorithm with $h=0.01$, }
\label{fig3}
\end{figure}

\begin{figure}
\begin{center}
\includegraphics[scale=0.3]{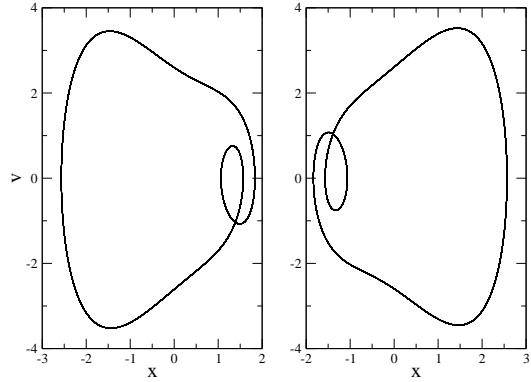}
\end{center}
\caption{Two periodic solutions of the forced Duffing oscillator with the same parameters $a=0.1$ and $b=3.5$ and different initial values of $x_0$ and $v_0$. The attractors have been plotted for $t>400$. The parameters of the polynomial algorithm: $\varepsilon=0.6, N=2$.
}
\label{fig4}
\end{figure}

\begin{figure}
\begin{center}
\includegraphics[scale=0.3]{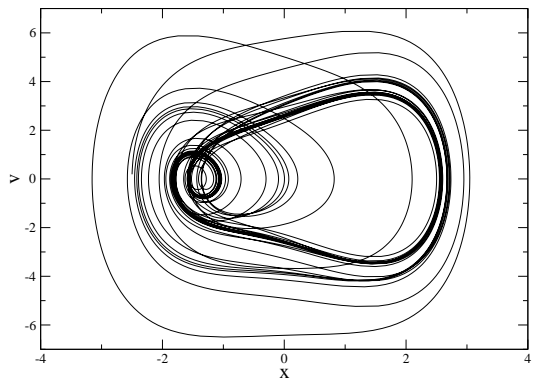}
\end{center}
\caption{The entire trajectory of the forced Duffing oscillator with the same parameters 
as in Fig.~\ref{fig4}  and $x_0=-1.5$ and $v_0=0$.}
\label{fig5}
\end{figure}
\begin{figure}[h]
\begin{center}
\includegraphics[scale=0.3]{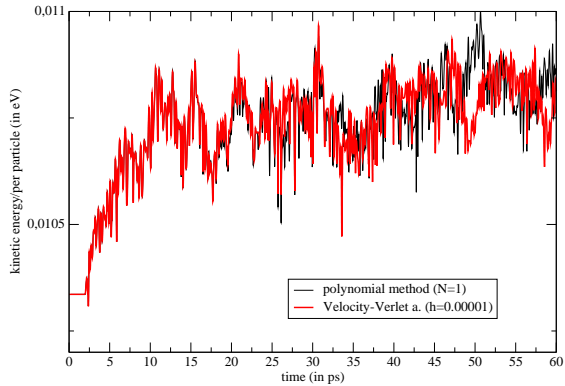}
\end{center}
\caption{Data representing 60ps of the kinetic energy per particle for the 2D Lennard-Jones fluid consisting of 500 particles. The chosen step size $h=0.00001$ in the case of the Verlet algorithm whereas $\varepsilon=0.0001$ and $N=1$ in the case of the polynomial method.}
\label{fig6}
\end{figure}

\section{Some features of the algorithm}
While performing numerical integrating motion equation one always 
is fighting for the numerical accuracy. In classical finite-difference methods like Verlet algorithm, leapfrog algorithm or Runge-Kutta algorithm this is connected with the chosen size $h$ of the time step. However, the smaller step size the larger cumulated round-off error  because more time steps are necessary to cover the given time interval. Thus, one should use a numerical method using the smaller number of steps (the larger value of $h$) without loss in numerical accuracy. The advantage of our method is already evident in Fig.~\ref{fig2}, where three solutions of the forced oscillator equation 
\begin{equation}
x^{\prime \prime}=-x+\cos(\omega t), \,\,\, x(0)=1,\,\,\,v(0)=0,
\end{equation}
\noindent
have been plotted,
the exact one represented by the equation 

\begin{equation}
x(t)=\frac{1}{1-\omega^2}( \cos(\omega t) - \omega^2 \cos(t)),  
\end{equation}

\noindent
and two numerical approximations represented by the Velocity-Verlet algorithm 
with the step size $h=0.01$ and our polynomial $x_N$ of degree $N=5$ in formal variable $\lambda$. In the case of the polynomial method there have been  plotted, in the figure, only the dots representing the points, where the condition Eq.~(\ref{r_TOL}) fails for a given accuracy $\varepsilon=0.01$. They are the only points where the numerical round-off errors contribute to the approximate solution. The remaining points (in between), which have not been plotted,  do not contribute to round-off errors cumulation. One always can recalculate them from the exact expression for the polynomial representation of $x(t)$.

The following advantage of our algorithm could be relatively shorter total calculation time than in any numerically stable finite-difference method in the limit of small values of $h$.
In Fig.~\ref{fig3}, we have presented calculation time dependence of the Velocity-Verlet algorithm on the value of $h$  and our polynomial algorithm
on the given accuracy $\varepsilon=h$. The results in the figure have been obtained from the programs calculating deviations of the approximate solutions from the exact one. The numerical errors arising from the assumed 
value of $\varepsilon$ can be by a few orders smaller than in classic finite-difference methods. This feature has already been discussed in our paper \cite{l_DudekNadzieja}, where we compared various numerical algorithms with respect to their numerical accuracy.

The next feature of the presented algorithm is that it applies also for strongly non-linear motion equations. In particular, in Fig.\ref{fig4}, we have presented two different attractors of the forced Duffing oscillator (the parameters have been taken from the Fig.~2.20 in a book by Holden \cite{l_Holden}) 
\begin{equation}
x^{\prime \prime}+a x^{\prime} + x^3 = b \cos(t), 
\end{equation}
\noindent 
with the same values of $a=0.1$ and $b=3.5$ but different initial conditions. 
In Fig.~\ref{fig5}, there has been presented entire trajectory starting from the initial condition and leading to one of the attractors.

One can use our method also for chaotic solutions of the oscillator. However, we do not discuss this possibility in this paper.

The polynomial $x_2(t)$, in the case of the Duffing oscillator, is represented by the following formulae:
\begin{eqnarray}
x_2(\tau) &= & x_0+\tau\,v_0 + 
\lambda\,[-\frac{1}{2}\,\tau^2\,v_0\,a-\frac{1}{2}\,x_0\,\tau^4\,v_0^2-\cos(t_0+\tau)\,b
-\frac{1}{2}\,x_0^3\,\tau^2	\nonumber\\
&& -\frac{1}{2}\,x_0^2\,\tau^3\,v_0+\cos(t_0)\,b-\frac{3}{10}\,\tau^5\,v_0^3-\tau\,b\,\sin(t_0)] \nonumber \\
&&+\lambda^2\,[24\,x_0\,\sin(t_0+\tau)\,b\,v_0+ 72\,\sin(t_0+\tau)\,\tau\,b\,v_0^2+\frac{11}{20}\,x_0^3\,\tau^6\,v_0^2
\nonumber \\
&&-12\,x_0\,\cos(t_0)\,\tau\,b\,v_0-b\,\sin(t_0)\,a+ \frac{1}{2}\,x_0^2\,\tau^3\,b\,\sin(t_0)\nonumber \\
&&-108\,\cos(t_0)\,b\,v_0^2 -3\,x_0^2\,\tau\,b\,\sin(t_0)-12\,x_0\,\tau\,\cos(t_0+\tau)\,b\,v_0 \nonumber\\
&&-\frac{3}{2}\,x_0^2\,\cos(t_0)\,\tau^2\,b +\frac{1}{6}\,x_0^3\,\tau^3\,a \nonumber \\ 
&&+108\,\cos(t_0+\tau)\,b\,v_0^2+36\,\tau\,b\,v_0^2\,\sin(t_0)-3\,x_0^2\,\cos(t_0+\tau)\,b \nonumber \\
&&+\frac{1}{8}\,x_0^5\,\tau^4-2\,x_0\,\cos(t_0)\,\tau^3\,b\,v_0+\frac{7}{20}\,\tau^6\,v_0^3\,a \nonumber \\
&&+\frac{9}{40}\,x_0\,\tau^8\,v_0^4-\frac{3}{2}\,\cos(t_0)\,\tau^4\,b\,v_0^2+\frac{1}{2}\,\tau^2\,b\,\sin(t_0)\,a \nonumber \\
&&-24\,x_0\,b\,v_0\,\sin(t_0)+\frac{53}{140}\,x_0^2\,\tau^7\,v_0^3+ \sin(t_0+\tau)\,b\,a \nonumber \\
&&+\frac{9}{10}\,\tau^5\,b\,v_0^2\,\sin(t_0)+\frac{1}{6}\,\tau^3\,v_0\,a^2+x_0\,\tau^4\,b\,v_0\,\sin(t_0)+ \nonumber\\
&&\frac{2}{5}\,x_0\,\tau^5\,v_0^2\,a-18\,\tau^2\,\cos(t_0+\tau)\,b\,v_0^2+\frac{3}{40}\,\tau^9\,v_0^5 \nonumber \\
&&+\frac{1}{4}\,x_0^2\,\tau^4\,v_0\,a+3\,x_0^2\,\cos(t_0)\,b- \cos(t_0)\,\tau\,b\,a+\frac{3}{8}\,x_0^4\,\tau^5\,v_0]  \nonumber \\
\end{eqnarray}
\noindent
and in this case the accepted approximated solution should satisfy the inequality $\vert \varphi_2(\tau) \vert < \varepsilon $ for a given $\varepsilon$, where  $\varphi_2(\tau)$ is the coefficient of $\lambda^2$ (we set the formal parameter $\lambda =1$ after all).

In our previous paper \cite{l_DudekNadzieja} we have shown that our method could be used also for molecular-dynamics simulations of large number of particles.
To this end, we have simulated barometric formula in the case of the ideal gas of $1000$ molecules in the gravitational field and the gas was contacted Nos\'e-Hoover thermostat \cite{l_Nose}, \cite{l_Hoover}.

In all mentioned by us cases, till now, the series expansion of the force (Eq.~(\ref{r_3})) consisted of a finite number of terms. The question arises, could the method be extended to a more general case, where the number of terms is infinite?  In order to show the possibility we have considered 
2D Lennard-Jones  fluid represented by a system of $N$ particles interacting with  Lennard-Jones 
 potential energy, 

\begin{equation}
U(r_{ij})=4 \varepsilon \left [ (\frac{\sigma}{r_{ij}})^{12} -(\frac{\sigma}{r_{ij}} )^6 \right ].
\end{equation}

\noindent
Then,  the force experienced by the particle $i$ from another particle $j$ being a distance $r_{ij}$ away is repesented by the following formula

\begin{equation}
F_i(r_{ij})=-\frac{dU(r_{ij})}{dr_{ij}}=\frac{24 \varepsilon}{\sigma} \left [ 2 (\frac{\sigma}{r_{ij}})^{13}-(\frac{\sigma}{r_{ij}})^7 \right ].
\end{equation}

\noindent
In this case the series expansion in the neighborhood of $(x_0,v_0)$ (see Eq.~\ref{r_3}) leads to an infinite number of terms including the powers of  
$\sqrt{(x_{0i}-x_{0j})^2+(y_{0i}-y_{0j})^2}$.

In the case of the approximating polynomials of the order of $\lambda^1$ the numerical algorithm  is equivalent to the Velocity-Verlet algorithm and it 
is represented by the following set of equations:

\begin{equation}
\overrightarrow{r_i(t)}=\overrightarrow{r_{i0}}+\overrightarrow{v_{i0}}\,{\tau}+ \frac{1}{m}(\overrightarrow{F_i}+ \overrightarrow{F_{iV}} \frac{\tau}{3})\frac{\tau^2}{2} ,
\label{r_poloz}
\end{equation}

\begin{equation}
\overrightarrow{v_i(t)}=\overrightarrow{v_{i0}}+\frac{1}{m}{\tau}(\overrightarrow{F_i} +\overrightarrow{F_{iV}} \frac{\tau}{2}) ,
\label{r_predkosc}
\end{equation}

\noindent 
where 

\begin{equation}
\overrightarrow{F_i} =(\overrightarrow{r_{i0}}-\overrightarrow{r_{j0}}) \vert{\overrightarrow{F_i}}\vert ,
\end{equation}

\begin{equation}
\overrightarrow{F_{iV}} =(\overrightarrow{v_{i0}}-\overrightarrow{v_{j0}}) \vert{\overrightarrow{F_i}}\vert \,.
\end{equation}

\noindent
and $\overrightarrow{r_{i0}}$ and $\overrightarrow{v_{i0}}$ are the initial location and velocity of the particle $i$, respectively. 

The accuracy control parameter $\varepsilon$ should satisfy the condition

\begin{equation}
\vert \frac{1}{m}(\overrightarrow{F_i}+ \overrightarrow{F_{iV}} \frac{\tau}{3})\frac{\tau^2}{2} \vert < \varepsilon.
\label{r_kondVerlet}
\end{equation}

\noindent
The generalization of the algorithm to the case of the polynomial approximation of the order $\lambda^2$ becomes much more complex and it is not presented in this paper.
However, already the results obtained in the linear approximation (in formal parameter $\lambda$) become promising. In Fig.~\ref{fig6}, there has been plotted the kinetic energy (per particle) of 500 particles representing 2D Lennard-Jones fluid versus time in the case of the Velocity-Verlet algorithm and our polynomial approximation (Eqs.~\ref{r_poloz}-\ref{r_predkosc}), linear in $\lambda$. 
In this case, the total  time $t_P$ used by the polynomial method  was of the same order as the total time $t_V$ of the Verlet method ($t_P=1.5\, t_V$).
In order to preserve the given numerical accuracy $\varepsilon$ the polynomial algorithm was runnining according to the following steps:  

\begin{enumerate}
\item[(i)] start with $\tau=h_0$, where $h_0=0.001$
\item[(ii)] if Eq.~(\ref{r_kondVerlet}) fails then change the value of $\tau$ by some factor, e.g. 
 $\tau=\tau/10$
\item[(iii)] calculate the values of the polynomials $\overrightarrow{r_i}$ and $\overrightarrow{v_i}$
\item[(iv)] goto (i).
\end{enumerate}

\noindent
In the considered example, the time intervals $\tau$ used during  the entire simulation run were distributed as follows:

\begin{eqnarray}
\tau = 10^{-3} & = & 0\% \nonumber\\
\tau  = 10^{-4} & = & 49\% \nonumber\\
\tau = 10^{-5} & = & 48.7\% \nonumber\\
\tau =  10^{-6} & = & 2\% .\nonumber\\
\end{eqnarray}

\noindent
The total calculation time strongly depends on the value of $\varepsilon$ and 
the higher order of the approximating polynomial (in the formal variable $\lambda$) makes possible larger values of $\tau$.

\section{Conclusions}
We have discussed possible numerical advantages of integrating motion equations  with the help of  the  recently published  algorithm for solving the initial-value problem for the ordinary differential equations \cite{l_DudekNadzieja}. Contrary to the traditional finite-difference methods, representing truncated series expantion of the solution of the equation of motion under consideration, the method is not discrete in time. This makes possible that for the large class of problems in physics the algorithm could be faster and more accurate than traditional finite-difference schemes. The particular example of the 2D Lennard-Jones fluid, which has been discussed above, suggests that  the method could be applied to the many-body problems 
~\\
~\\
~\\
\noindent
{\bf Acknowledgments}\\
We are indebted for discussion with Prof. K. Wojciechowski and Prof. A.C. Bra\'nka. We also thank Prof. W.G. Hoover  for his comments on the algorithm and the suggestion of some numerical tests.

\end{document}